\documentclass[aps,prb,preprint,superscriptaddress,amsmath,amssymb,floatfix,raggedbottom]{revtex4-1}

\usepackage{graphicx}% Include figure files
\usepackage{dcolumn}% Align table columns on decimal point
\usepackage{bm}% bold math
\usepackage[utf8]{inputenc}
\usepackage[T1]{fontenc}
\usepackage{mathptmx}
\usepackage{etoolbox}
\usepackage{color}
\usepackage{soul}
\usepackage{comment}
\usepackage{xr}         % or xr-hyper if using hyperref

\begin{document}

\title{Interaction between point defects and vertical inversion domain walls in wurtzite AlN}
% Force line breaks with \\
\author{L. Naudin}
 \affiliation{ 
Université de Toulouse, CNRS, CEMES, Toulouse, France %\\This line break forced with \textbackslash\textbackslash
}%

\author{C. Paillard}%
\affiliation{%
Smart Ferroic Materials Center, Physics Department and Institute for Nanoscience and Engineering, University of Arkansas, Fayetteville, Arkansas 72701, USA%\\This line break forced% with \\
}%
\affiliation{Université Paris-Saclay, CentraleSupélec, CNRS, Laboratoire SPMS, 91190 Gif-sur-Yvette, France}

\author{L. Bellaiche}
\affiliation{%
Smart Ferroic Materials Center, Physics Department and Institute for Nanoscience and Engineering, University of Arkansas, Fayetteville, Arkansas 72701, USA%\\This line break forced% with \\
}%
 \affiliation{Department of Materials Science and Engineering, Tel Aviv University, Ramat Aviv, Tel Aviv 6997801, Israel}%Lines break automatically or can be forced with \\

\author{L. Calmels}
 \affiliation{ 
CEMES-CNRS, Université de Toulouse, 29 rue Jeanne Marvig, 31055 Toulouse%\\This line break forced with \textbackslash\textbackslash
}%

\author{R. Arras}
 \affiliation{ 
CEMES-CNRS, Université de Toulouse, 29 rue Jeanne Marvig, 31055 Toulouse%\\This line break forced with \textbackslash\textbackslash
}%

 \email{loris.naudin@cemes.fr; remi.arras@cemes.fr}

\newcommand{\remi}[1]{\textcolor{magenta}{{#1}}} 
\newcommand{\loris}[1]{\textcolor{cyan}{{#1}}} 

\date{\today}% It is always \today, today,

\begin{abstract}
Alloyed aluminium nitride compounds constitute a promising class of ferroelectric materials due to their high remanent electric polarizations, large band gaps and structural compatibility with a growth on Si substrates. Such materials nonetheless possess large coercive fields and polarization-switching mechanisms are still debated. We performed first-principles calculations to investigate the stability of isolated point defects in the vicinity of a vertical inversion domain wall (DW). We found that all studied defects are energetically more stable at or near the DW. Depending on their nature, they can have the opposite effect on the displacement of the DW, which occurs during polarization switching. Finally, we discuss how likely the different defects may be responsible for leaking currents and degraded ferroelectric properties.
\end{abstract}

%119

\maketitle

%\section{Introduction}

Wurtzite AlN is a semiconductor with a wide-band-gap energy $E_\textrm{g}\sim 6.0$–6.2~eV that has recently attracted significant attention due to the discovery of parent compounds which form a new class of ferroelectric materials. When alloyed with elements such as Sc, B or Gd, AlN indeed exhibits a robust ferroelectric behavior~\cite{Fichtner:2019,Hayden:2021,Lee:2025} combining large remanent polarization, up to $ \sim 100~\mu$C~cm$^{-2}$), with high thermal~\cite{Islam:2021} and chemical stability.
 
AlN alloys have emerged as promising candidates for integration in next-generation electronic and optoelectronic devices due to their compatibility with the fabrication processes for Complementary Metal Oxide Semiconductor (CMOS) platforms~\cite{Wang:2023} as well as their exceptional structural and ferroelectric properties. Applications include ferroelectric random access memories (FeRAM)~\cite{Zhang:2024}, radiofrequency devices~\cite{Izhar:2025,Zhang:2025}, MEMS sensors and actuators~\cite{Abdelaal:2025} or electro-optical modulators~\cite{Wang:2025}.  

Nevertheless, ferroelectric nitrides possess significant shortcomings that prevent their use as a major constituent of future devices. In particular, their coercive field $E_\textrm{c}$ is close to the breakdown field~\cite{Zhang:2024}. Wang {\it et al.} demonstrated that it is possible to decrease the value of $E_\textrm{c}$ from 6~MV~cm$^{-2}$ to less than 4~MV~cm$^{-2}$ by increasing AlN doping from 15\% to 35\% Sc, respectively. However, such values of $E_\textrm{c}$ remain rather high and are obtained to the detriment of lower band gaps and saturation polarizations (P$_\textrm{S}$). Other strategies to reduce the coercive field have then been proposed, including strain~\cite{Yassine:2022,Liu:2025}, electrode~\cite{Zhang2025_2} or interfacial engineering~\cite{Joo:2024} and co-alloying~\cite{Saha:2025,Bradford:2025}. Controlling point defects like vacancies, interstitials, or substitutions could also be a viable route to reduce coercive fields~\cite{Lee:2024}. In ferroelectric materials, domain walls (DWs) play a central role in determining macroscopic properties such as coercive fields, switching kinetics, dielectric response~\cite{Quan:2015} and fatigue behavior~\cite{Sunil:2025}. 

With the goal of further decreasing the magnitude of ferroelectric-nitrides coercive fields, it is necessary to gain a better understanding of the fine mechanisms responsible for their electric polarization switching. For symmetry reasons, wurtzite nitride ferroelectrics have an electric polarization oriented along the hexagonal [0001] direction only~\cite{Fichtner:2025}, thus implying that DWs in wurtzite nitrides are all 180° oriented DWs, disclosing the possibility to have complex DW patterns as found in perovskite structures. To explain polarization switching in Al$_{1-x}$Sc$_x$N alloys, the first models proposed in the literature involved the formation of an extended non-polar intermediate phase, with a hexagonal layered phase~\cite{Fichtner:2019} or a $\beta$-BeO structure~\cite{Calderon:2023,Lee:2024_2}. Recently, the formation of atomically-sharp DWs has been experimentally evidenced~\cite{Wolff:2025,Wang:2025_2} and it has been shown that polarization switching occurs by reversing the polarization state of atomic columns at the domain wall~\cite{Behrendt:2026,Huang:2026}. Recent theoretical studies~\cite{Behrendt:2026,Huang:2026} have shown that DW displacement occurs stochastically chain-by-chain, giving rise to rough DWs. This mechanism could also be responsible for the so-called proximity ferroelectricity effect in wurtzite ferroelectrics~\cite{Skidmore:2025, Eliseev:2025}.

Various sharp domain wall configurations have been identified, which are vertical~\cite{Umar:2021, Wolff:2025} (i.e., parallel to the polarization axis), horizontal~\cite{Kato:2024} (i.e., perpendicular), or inclined~\cite{Zhang:2022,Wolff:2025}. Point defects such as vacancies, antisites, and impurities are ubiquitous in AlN thin films grown by sputtering or metalorganic chemical vapor deposition (MOCVD)~\cite{Zhang:2024}, therefore a better understanding of their formation and interaction with DWs is warranted. In fact, the structure and mobility of DWs are often strongly affected by the presence of point defects, which can act as pinning centers~\cite{Zhao:2026, Lee:2024, Geng:2020, Paruch:2013}, charge traps~\cite{Vega:2025}, or nucleation sites~\cite{Gao:2011} for polarization switching. Engineering of point defects in AlN can moreover have applications in resistive switching memories~\cite{Yang:2022, Guido:2023, Guo:2024} or quantum information~\cite{Varley:2016, Bishop:2020, Xue:2020, Wang:2023_2, Czelej:2024, Zhu:2024, Zhu:2025}.

In this paper, we use first-principles calculations based on density functional theory to investigate the atomic structure, thermodynamical stability, and electronic properties of vertical domain walls combined with the presence of single point defects in wurtzite AlN. We choose to focus our study on vertical DWs only, as they are widely reported in the scientific literature~\cite{Wolff:2025}, possess a simpler structure and the advantage of being electrically neutral. We considered nine different, neutral or charged, point defects that are either among the most commonly found in this system or of technological relevance; these defects consist in two atomic vacancies V$_\textrm{N}$ and V$_\textrm{Al}$ and seven substitutions X$_\textrm{N}$ (with X = O or C) and X$_\textrm{Al}$  (X= B, Si, Sc, Y, La). 

%657 mots

%\section{Details of calculations}

First-principles calculations were performed using the Vienna \textit{ab initio} Simulation Package (VASP)~\cite{Kresse:1996a, Kresse:1996b} and the projector augmented-wave (PAW) method~\cite{Blochl:1994, Kresse:1999}. The exchange–correlation energy was treated within the generalized gradient approximation (GGA) using the parameterization of PBEsol~\cite{Perdew:2008,Perdew:2009}. More details are given in Sec.~I of the supplementary-material file~$^\star$. 

\begin{figure}[h]
    \centering
    \includegraphics[width=1.0\linewidth]{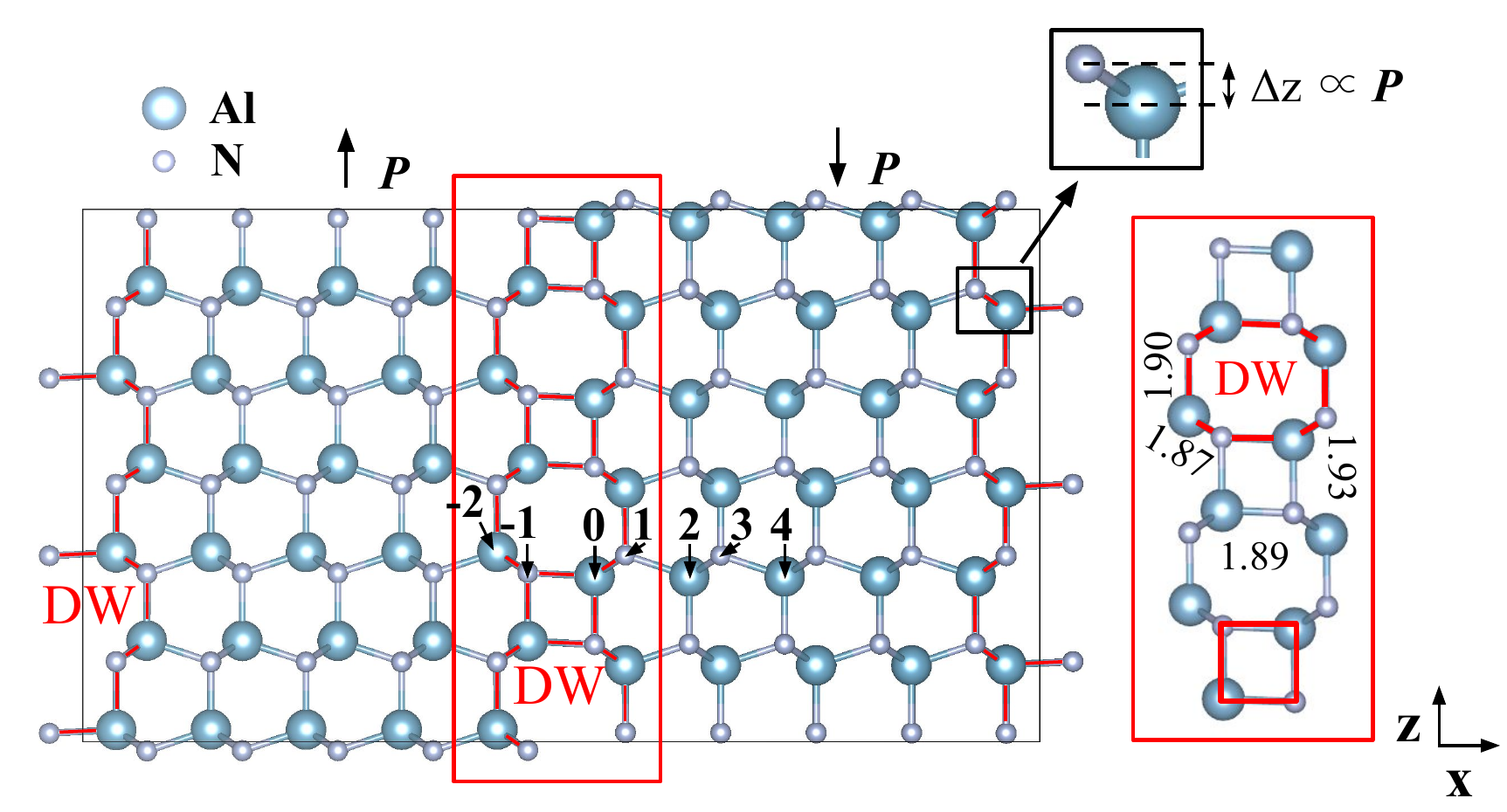}
    \caption{(left) Atomic structure of the AlN supercell with the two equivalent DWs highlighted in red. The numbers are used to label the $(10\bar{1}0)$ atomic layers parallel to the DW. (right) Details of the Al-N chemical-bond lengths at the DW (given in \AA).  \label{fig:figure1_DW}}
\end{figure}

We performed our calculations using either a $4\times 4 \times 3$  bulk supercell or, as displayed in Fig.~\ref{fig:figure1_DW}, a $10 \times 4 \times 3$ supercell of the wurtzite structure of AlN containing two equivalent DWs with a thickness of two atomic layers perpendicular to the $x$ axis of $[10\bar{1}0]$ hexagonal direction. These DWs have the same structure as proposed in previous studies~\cite{Zhu:2022, Wolff:2025} and they separate two ferroelectric domains of opposite electric polarization $\bm{P}$ and with initially the same thickness. The electric polarization is aligned along the $z$ axis (hexagonal direction [0001]) and emerges as an off-centering displacement of the Al and N ions from the center of their tetrahedra atomic sites, which can be characterized by the rumpling parameter $\Delta z = z(\textrm{Al})-z(\textrm{N})$, that we found equal to 0.59~\AA{} in the bulk (see Fig. 1). Each atom possesses four first neighbors (1NN), three being in the tetrahedron basal plane and one being out-of-plane.

%230 mots

%\section{Results}

%\paragraph{DW without point defects:}
At the DW (atomic layers No $-2$ to $1$), the $\Delta z$ parameter is on average equal to 0~\AA{}. It is locally of 0.42~\AA{} if we instead consider only one half the DW (layers No $0$ and $1$), which consists in a local reduction by $\sim$ 29\% compared to the bulk value. The presence of the DW also induces structural distortions compared to the bulk compound, with an alternating increase/decrease of out-of-plane bonds by 1.42\%/$-0.32$\% (from 1.903~\AA{} to  1.930~\AA{}/1.897~\AA{}) and a shortening of basal in-plane bonds by 0.85\% on average (from 1.891~\AA{} to 1.875~\AA{}) as shown in Fig.~\ref{fig:figure1_DW}. In addition, the DW induces a slight decrease by 0.20~eV of the band-gap energy at the Fermi level (Fig.~S1~$^\star$), in agreement with the calculations reported in Ref.~\cite{Hwang:2024}. Its formation energy was calculated to be 32.90~meV~\AA$^{-2}$. 

%\paragraph{Stability of point defects versus position from DW:} 
We now discuss the relative energetic stability of these point defects with respect to the distance from the DW. As labeled in Fig.~\ref{fig:figure1_DW}, five atomic layers $d$ (0, 1, 2, 3 and 4) have been considered to locate the point defects, at different distances from the DW. 

\begin{figure}[h]
    \centering
    \includegraphics[width=0.5\linewidth]{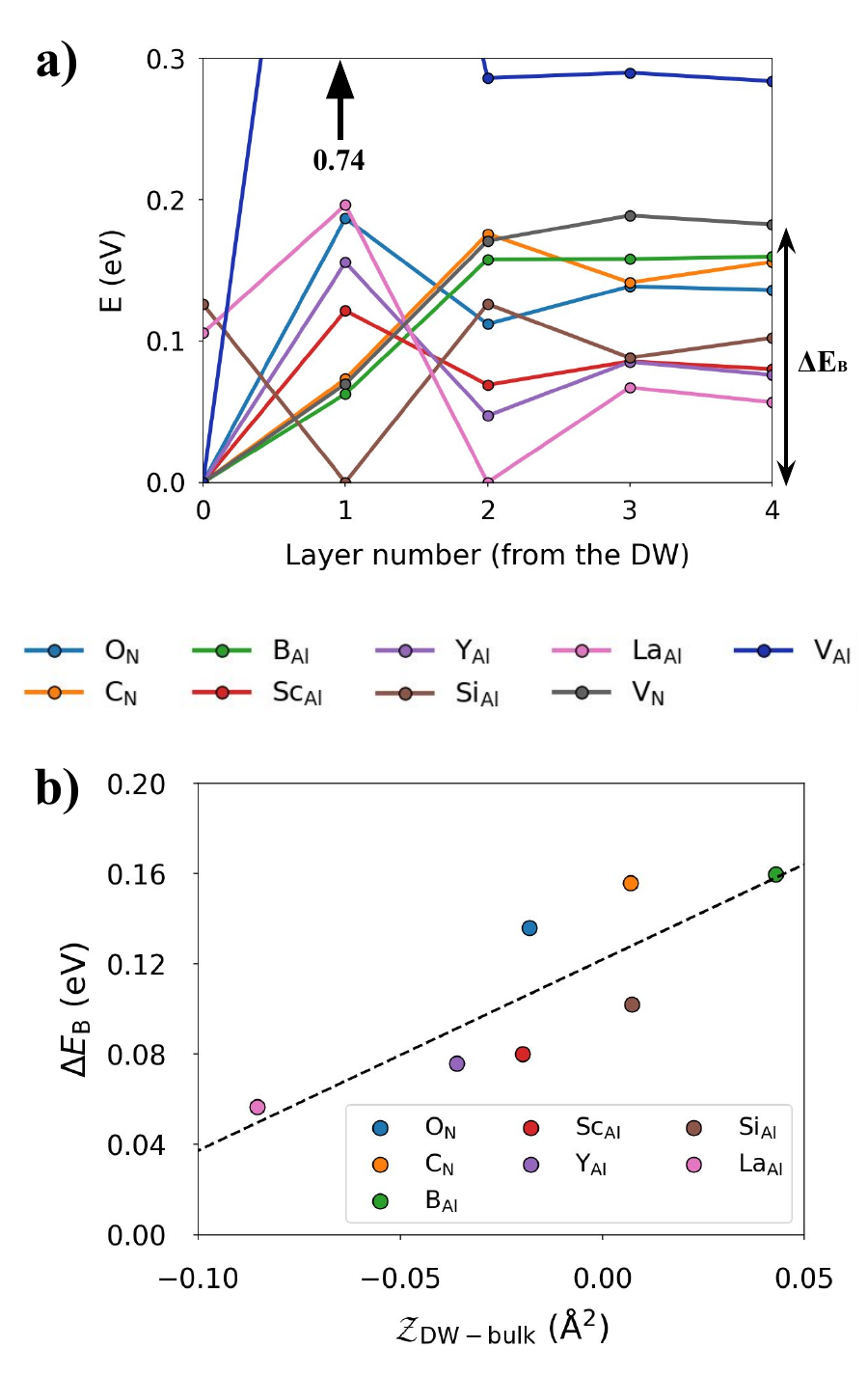}
    \caption{a) Relative energy $E$ calculated for each defect with respect to their distance to the DW. The origin of the energies is chosen according to the most stable position for each defect. b) Energy gain $\Delta E_\textrm{B}$ at the DW with respect to the $\mathcal{Z}_\textrm{DW-bulk}$ parameter defined in Eq.~\ref{eq:Eq2}.}
    \label{fig:fig2}
\end{figure}

The variation of the calculated relative total energies $E$ for the different defect locations is plotted in Fig.~\ref{fig:fig2}(a). Except for La$_\textrm{Al}$ substitutions, all defects are found to form preferentially at the DW, similarly to the results reported in ferroelectric perovskites~\cite{Chandrasekaran:2013, Paillard:2017}. Vacancies (V$_\textrm{Al}$, V$_\textrm{N}$) have the largest energetic stabilization out of all point defects. The stability of point defects at the DW can be qualitatively attributed to two different effects: (i) out-of-plane and basal in-plane bonds are significantly modified at the DW in comparison to their bulk values and these structural distortions may facilitate the formation of point defects at the DW by releasing elastic strain, (ii) because the electric polarization is locally decreased at the DW, the additional energy cost resulting from a weakening of the ferroelectricity is reduced if the point defect is located on the DW. 

In Fig.~S2, we reported local structural and energetic variations induced by the presence of the different point defects in bulk AlN. Depending on the considered defect, substantial variations of the local strain are induced, mostly associated with changes of the volumes formed by N or Al tetrahedra (Fig.~S2b). Nonetheless, no significant gain is expected from a localization at the DW as these volumes of tetrahedra are almost equal at the DW or far from it, which allows to discard this effect (i) as the main cause for the stabilization of defect at the DW. On the contrary, we can notice that the local $\Delta z$ associated with scenario (ii) could be a more natural parameter to explain the accommodation of the defect as it is decreased by $\sim 29$\% between bulk and DW (Fig.~S2c) and it can be considered in first approximation as proportional to the electric polarization $P$. 

We tried to explain the energy gain obtained when a defect is located at the DW instead of in the bulk by considering a Landau model similar to the one proposed in Ref.~\cite{Paillard:2017}, which leads to the expression: 

\begin{equation}
\label{eq:Eq1}
\Delta E_\textrm{B} = E^\textrm{d=0}_\textrm{X}-E^{d = 4}_\textrm{X} \simeq \alpha\mathcal{Z}_\textrm{DW-bulk} + \beta
\end{equation}

with 

\begin{equation}
\label{eq:Eq2}
\mathcal{Z}_\textrm{DW-bulk} \simeq \Delta z_\textrm{AlN}^\textrm{d = 0}\delta(\Delta z)_\textrm{X}^\textrm{d = 0} - \delta z_\textrm{AlN}^\textrm{d = 4}\delta(\Delta z)_\textrm{X}^\textrm{d = 4}    
\end{equation}
In Eq.~\ref{eq:Eq2}, we consider that $\Delta z$ is proportional to $P$ and $\delta (\Delta z)_\textrm{X}$ to $\Delta P$, i.e. $\delta (\Delta z)_\textrm{X} = \Delta z_\textrm{X}-\Delta z_\textrm{AlN}$. $\Delta z_\textrm{AlN}$ is the local $\Delta z$ parameter calculated in pure bulk AlN and $\Delta z_\textrm{X}$ the $\Delta z$ calculated by averaging on Al or N atoms first-neighbors of the defect X. $E^{d = 4}_\textrm{X}$ is the defect energy calculated at the position $d = 4$, i.e., far from  the DW, while $E^{d = 0}_\textrm{X}$ is the energy of the defect at the DW. More details are given in Sec.~SII.b~$^\star$.

The energy difference $\Delta \mathrm{E_B}$ plotted in Fig.~\ref{fig:fig2}(b) for the different substitutions as a function of $\mathcal{Z}_\textrm{DW-bulk}$ displays a good linear correlation between these two parameters (coefficient correlation of $r = -0.83$). We can see a tendency for small atoms (B, C, O) to be more easily stabilized at the DW, while larger atoms (Sc, Y) are expected to show a more random distribution between bulk and DW. These results are in agreement with the variations of on-site $\Delta z$ presented in Fig.~S2c. However, considering also the significant variations of above-site first-neighbors rumplings (Fig.~S2d), i.e. the non-locality of the defect-induced polar variation, it is important to keep in mind that using only the variation of on-site $\Delta z$ parameter may be a rough approximation to explain the variation of $\Delta E_\textrm{B}$.

Up-to-now, for the sake of simplicity, we focused our paper on the effects of neutral point defects. Excluding substitutions with isoelectronic atoms, the formation of charge defects is highly probable and will depend on the electronic chemical potential associated to experimental conditions~\cite{Lee:2024, Osetsky:2022, Aleksandrov:2020}. If electrostatic interaction between charge defects and the electrically-neutral DW, where $P$ is locally reduced, should be minimal, the charge state $q$ will also influence the ionic radius, the chemical bond length and the resulting local strain (Fig.~S3~$^\star$). In Fig.~S4~$^\star$, we reported the calculated energetic stabilities of charged point defects compared to their neutral counterparts. For every charged defect, the DW remains the more stable position and the stability variations of O$_\textrm{N}^{+1}$ and Si$_\textrm{Al}^{+1}$ do not show significant changes in comparison with their neutral counterparts. The main changes are obtained for the vacancies: for V$_\textrm{Al}^{-3}$, $\Delta E_\textrm{B}$ increases by $\sim 0.37$~eV (i.e., $+130$\%) compared to V$_\textrm{Al}^{0}$; singly-charged nitrogen vacancies V$_\textrm{N}^{+1 \textrm{,} -1}$ do not change noticeably $\Delta E_\textrm{B}$ whereas V$_\textrm{N}^{+2}$ increases $\Delta E_\textrm{B}$ by $\sim 0.26$~eV ($+141$\%) and V$_\textrm{N}^{+3}$ by $\sim 0.31$~eV ($+167$\%).

\begin{figure}[h]
    \centering
    \includegraphics[width=0.5\linewidth]{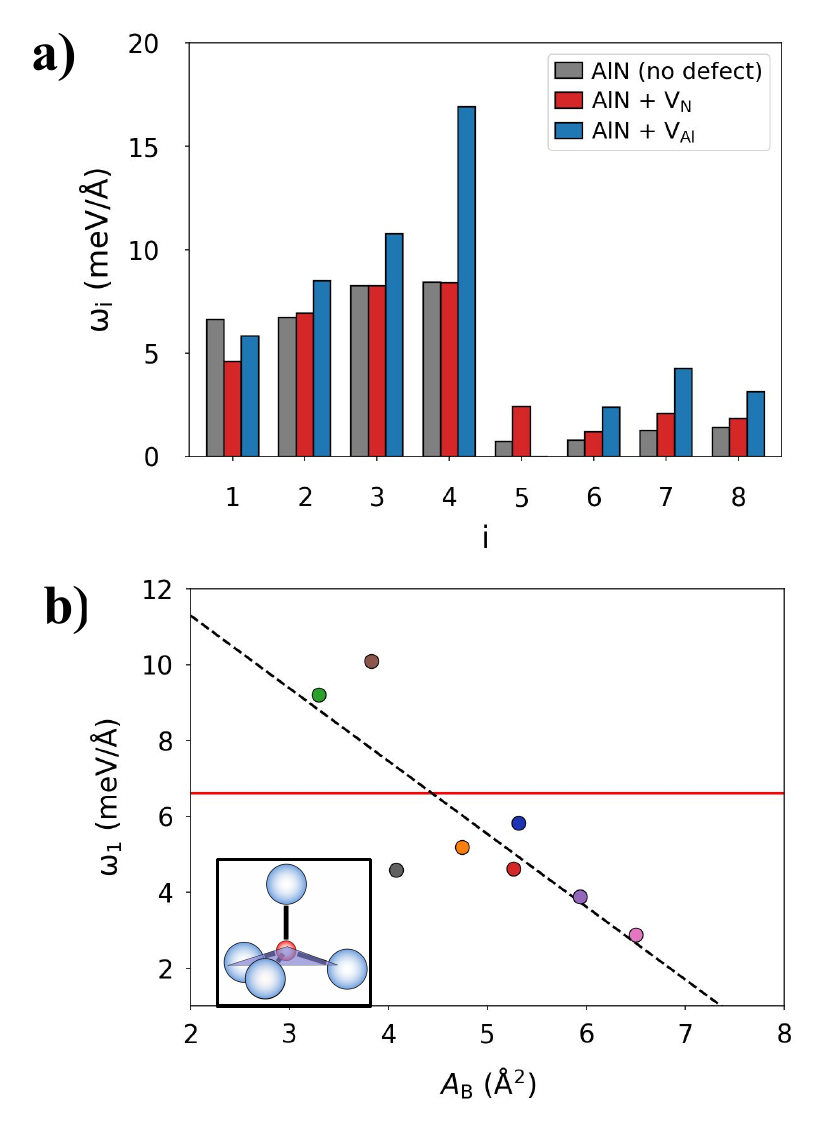}
    \caption{a) Energy barriers $\omega_i$ ($i=1,...8$) associated with the variation of the polar state of (001)-oriented atomic columns during the displacement of the DW by one unit cell along the $[10\bar{1}0]$ direction (see Fig.~S5), with and without the presence of one atomic vacancy.   b) First energy barrier $\omega_1$ as a function of the tetrahedra basal area $A_\textrm{B}$. \label{fig:fig3}}
\end{figure}

%\paragraph{Displacements of DW:} 
To understand the effects of point defects on the processes responsible for the switching of the electric polarization, we performed nudged elastic bands (NEB) calculations~\cite{Mills:1995} to calculate the energy pathway during the displacement of one DW along the $x$ axis. According to our calculations (Fig.~S5~$^\star$) and in agreement with Ref.~\cite{Behrendt:2026}, the displacement of the DW occurs through successive changes of polar states of [0001]-aligned atomic columns. As shown in Fig.~\ref{fig:fig3}(a) and in Fig.~S5(c)~$^\star$, considering our supercell dimensions containing four of these columns, the full displacement of the DW by one atomic layer thus goes through seven meta-stable states and overcomes eight barrier energies $\omega_i$.

When the structure has no point defects, all these barrier energies are close, which tends to show that the modifications of the polar state of the different atomic chains are quasi independent. If a V$_\textrm{N}$ defect is introduced, the energy barrier to switch the atomic chain containing the defect first is lowered by 2.02~meV~\AA$^{-2}$; on the contrary, the barrier energies associated with all other subsequent chains increase. The presence of such a defect could therefore facilitate initiating the displacement of a single atomic chain, in agreement with the results given in Ref.~\cite{Dryzhakov:2026}, but would overall have a pinning effect and randomize the polar states near the DW. When a N vacancy is present, the total energy barrier $\omega_\textrm{tot} = \sum_i \omega_i$ for the displacement of the DW by one atomic layer increases from $\sim 34.15$~meV~\AA$^{-2}$ for pure AlN to $\sim 35.73$~meV~\AA$^{-2}$ ($+4.63$\%). For V$_\textrm{Al}$, the pinning effect is predicted to be much stronger, with $\omega_\textrm{tot}$ increasing up to 51.76~meV~\AA$^{-2}$ ($+51.54$\%). 

Figure~\ref{fig:fig3}(b) shows the variations of energy barriers $\omega_1$ that must be overcome to change the polar state of the atomic column containing a single point defect, with respect to the basal triangle area $A_\textrm{B}$ defined by the surrounding Al or N first-neighbor atoms (see Fig.~2Sa). A strong correlation ($r=-0.82$) is obtained, corresponding to a decrease of $\omega_1$ with increasing $A_\textrm{B}$. This result shows that the ability to modify the polar state of an atomic column is linked to the structural hindrance with its neighboring atoms. Another consequence of the inclusion of an atomic defect is a modification of the $\Delta z$ parameter, as shown in Fig.~S6. The reduction of $\omega_1$ when $\Delta z$ decreases is consistent with the calculations on O substitutions and N vacancies reported in bulk Sc-doped AlN~\cite{Lee:2024} and in bulk LaN~\cite{Rowberg:2021}, which has been explained by the closer proximity to the non-polar intermediate phase which mediates the polarization switching. According to our calculations, we can see that only Si$_\textrm{Al}$ and B$_\textrm{Al}$ substitutions tend to increase $\omega_1$, by $\sim+53$ \% and $\sim+39$\%, respectively, in comparison with pure AlN. Regarding the B$_\textrm{Al}$ substitution, our result is consistent with the decrease of $A_\textrm{B}$ and $\Delta z$ parameters (Figs.~S2a and c) and in agreement with calculated increase of the electric polarization reported in Ref.~\cite{Hayden:2021}. Alloying of AlN with B atoms being known to help for the stabilization of ferroelectricity~\cite{Hayden:2021, Liu:2023, Zhang:2025_3}, our calculated increase of $\omega_1$ suggests that either other $\omega_i$ barrier energies are reduced, leading to a lower $\omega_\textrm{tot}$ global barrier, or that collective effects and disorder play a critical role in AlBN alloys. Finally, the substitution O$_\textrm{N}$ does not appear in Fig.~\ref{fig:fig3}(b) as it is not possible to modify the polar state of the chain containing this defect alone (see Fig.~S7a~$^\star$), indicating that such a defect is expected to induce a strong pinning of domain walls, in particular if their content is significantly high. This effect may be particularly detrimental when alloying with Sc atoms is used, as such chemical species have a higher affinity with oxygen atoms. In Fig.~S7b~$^\star$, considering the example of Si$_\textrm{Al}$ substitution, we highlight that changing the charge state, from $q = 0$ to $q = +1$, may also modify the first-displacement barrier height, as it affects both the induced structural distortions and the electronic structure. 

Finally, because calculating the full chain-by-chain reaction path to displace the DW for every defect was computationally too demanding, we instead computed the variation of energy barrier for a coherent, but normally energetically more costly, displacement of the DW for the most significant defects, this in order to have a complementary idea of the pinning effect (Fig.~S8~$^\star$). With this displacement mechanism, Sc$_\textrm{Al}$ substitution leads to a decrease of the total barrier, B$_\textrm{Al}$, on the contrary, induces an increase of $\omega_\textrm{tot}$, which remains in agreement with our previous results. Again, the neutral O$_\textrm{N}$ substitution is more detrimental for the displacement of the DW and does not allow for the stabilization of any intermediate metastable structure. The presence of a N vacancy is an exception as it induces a decrease of $\omega_\textrm{tot}$, which is different from its predicted increase with the ``chain-by-chain'' and energetically more effective mechanism; the coherent displacement of the DW with a V$_\textrm{N}$ defect is even found energetically more favorable, by 1.3\%, than the ``chain-by-chain'' one without defect. 

\begin{figure}[h]
    \centering
    \includegraphics[width=0.5\linewidth]{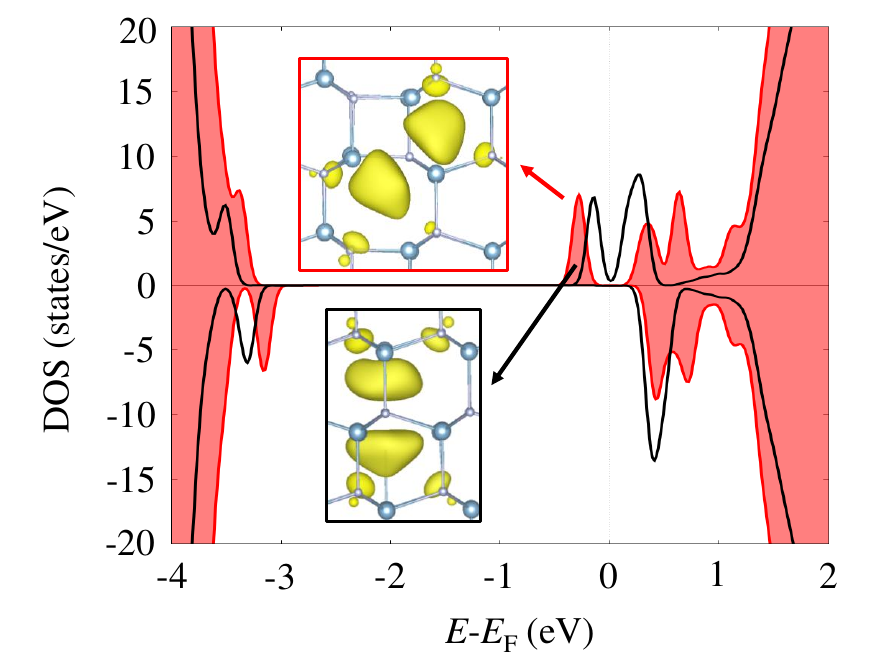}
    \caption{DOS calculated for a neutral VN located at the DW (red) and in bulk AlN (black) and charge densities corresponding to the localized defect states below the Fermi level. \label{fig:fig4}}
\end{figure}

%\paragraph{Effects on electronic properties:} 
Among the applications which would benefit from the creation and migration of point defects, resistive switching is promising for the development of neuromorphic computing. In Refs.~\cite{Guido:2023,Guo:2024}, the formation of conductive filament has been attributed to the migration of N vacancies; we thus chose to focus on this defect as an example. 

In Fig.~\ref{fig:fig4}, we display the density of states (DOS) calculated when a neutral V$_\textrm{N}$ defect is present at the DW. Spin-polarized defect states appear near the top of the valence bands and the bottom of the conduction bands. Just below the Fermi level, we have the presence of a peak of DOS formed by $s$ and $p$ states. When V$_\textrm{N}$ is located at the DW, we can observe a band gap of approximately 250~meV, which is closed when the vacancy is located in the bulk and is associated with a variation of the orbital symmetries of the occupied state. DOS for other defects are given for information in Fig.~S9 and S10 of the supplementary information file~$^\star$. V$_\textrm{N}$ induces the appearance of gap states both near the VBM and CBM, C$_\textrm{N}$ and V$_\textrm{Al}$ only near the VBM. O$_\textrm{N}$ and Si$_\textrm{Al}$ neutral substitutions get metallized with transfer of electrons in the CBM. Isoelectronic defects may only slightly decrease the band gap, without adding gap states, but can modify the orbital nature of the band edges, with added contributions of $3d$ orbitals at the CBM when Sc atoms are introduced, for example. 

%2080

%\section{Conclusion}

In conclusion, we performed a systematic study of the interaction between point defects and vertical inversion DW in AlN. We first showed that all point defects are stable at or near the DW. The better stability at the DW can be partly explained by the associated variation of cation-anion rumpling resulting in an energy gain due to the partial electric-polarization cancellation.
We also demonstrated that the presence of most of the point defects (V$_\textrm{N}$, V$_\textrm{Al}$, Sc$_\textrm{Al}$, Y$_\textrm{Al}$, C$_\textrm{N}$, and in particular La$_\textrm{Al}$) could make the initiation of the resistive switching easier, because of hindrance effect and a closer proximity with the non-polar hexagonal phase. Nonetheless, in overall, such defects can pin the DW or modify the switching mechanism, as shown for V$_\textrm{N}$ and V$_\textrm{Al}$. Other defects such as O$_\textrm{N}$ or B$_\textrm{Al}$ substitutions are found to be particularly detrimental for the polarization switching, at least if these defects are isolated, collective effects of the defects certainly leading to a totally different behavior~\cite{Lee:2024_2, Huang:2026}. We confirm that the presence of the DW slightly decreases the band gap energy and the additional presence of point defects which further reduce the band gap or introduce shallow states may induce leakage currents. While detrimental to preserve good ferroelectric properties, the induced electronic conductivity may have important implications for applications like resistive switching. The current study opens routes toward more advanced study to understand the presence and the role of complexes of defects and of a more or less large number of simultaneous substitutions, in particular when voluntarily alloying AlN to form a ferroelectric material, and for which disorder may play a critical role.

%267

\begin{acknowledgments}
This work was granted access to the HPC resources of CALMIP (Allocation No. 2024-2026/P19004) and CINES (Allocation AD010915807R1). Work at the University of Arkansas has been supported by a grant from the US Department of Energy under Award no. DE-SC0025479 and supported by the Arkansas High Performance Computing Center which is funded through multiple National Science Foundation grants and the Arkansas Economic Development Commission.          
\end{acknowledgments}

\nocite{*}
%\printbibliography% Produces the bibliography via BibTeX.

\section*{Supplementary materials: Interaction between point defects and vertical inversion domain walls in wurtzite AlN}

\graphicspath{ {figures/} }

\newcommand{\etal}{\textit{et al.} }

\setcounter{figure}{0}
\renewcommand{\figurename}{Fig.}
\renewcommand{\thefigure}{S\arabic{figure}}

\setcounter{figure}{0}
\renewcommand{\tablename}{Table}
\renewcommand{\thetable}{S\arabic{table}}

\renewcommand{\thesection}{S.\Roman{section}}

\renewcommand{\theequation}{S.\arabic{equation}}

\section{Details of calculations}

First-principles calculations were performed using the Vienna \textit{ab initio} Simulation Package (VASP)~\cite{Kresse:1996a,Kresse:1996b} and the projector augmented-wave (PAW) method~\cite{Blochl:1994,Kresse:1999} with a plane-wave cutoff energy of 600~eV. The exchange–correlation energy was treated within the generalized gradient approximation (GGA) using the PBEsol parametrization~\cite{Perdew:2008,Perdew:2009}.

We used two supercells for the simulations: a $4\times4\times3$ bulk  supercell and a $10\times4\times3$ supercell containing two equivalent vertical domain walls separating two domains of opposite electric polarization . The calculated equilibrium lattice parameters of the AlN unit cell are $a = b = 3.11$~\AA{} and $c = 4.98$~\AA, in agreement with experimental parameters reported in Refs.~\cite{Schulz:1977,Nilsson:2016}.

All structures were fully relaxed (lattice parameter relaxation included) until the Hellmann–Feynman forces on each atom were less than 0.01~eV~Å$^{-1}$.

Brillouin zone integrations were performed using the Monkhorst–Pack scheme~\cite{Monkhorst:1976}, with a $4\times4\times3$ and a $1\times4\times3$ $\bm{k}$-point grid chosen to ensure well-converged total energies and forces for the supercell. Static calculations were performed on the fully relaxed geometries using denser $\bm{k}$ meshes ($5\times5\times4$ and $1\times5\times4$).

\newpage

\section{Results}

\subsection{Density of states at the vertical domain walls without defects}

\begin{figure}[htb!]
    \centering
    \includegraphics[width=0.8\linewidth]{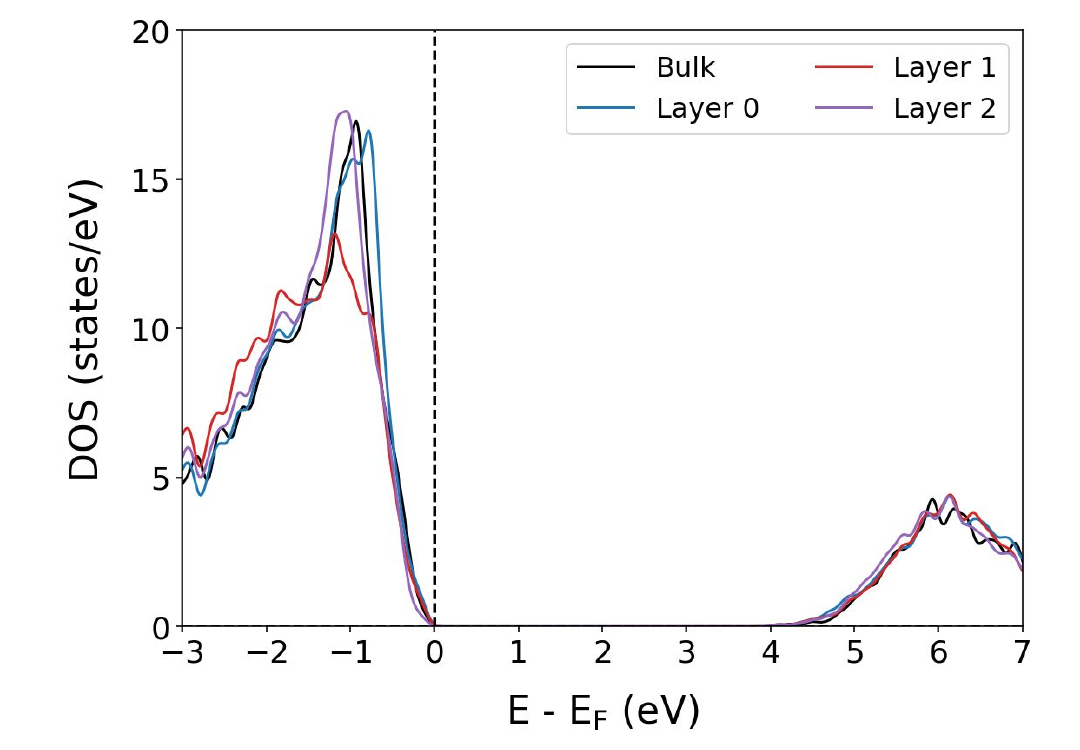}
    \caption{Layer-resolved DOS calculated for different atomic planes parallel to the domain wall (red curve) and compared with the DOS of bulk AlN compound (black curve). The labelling of the layers is given in Fig.~1 of the main paper.  }
    \label{fig:DOSBULKvsDW}
\end{figure}

\newpage 

\subsection{Interactions between the domain wall and the point defects}

\subsubsection{Variations of the atomic structure induced by point defects in bulk AlN:}

\begin{figure}[hb!]
    \centering
    \includegraphics[width=1.0\linewidth]{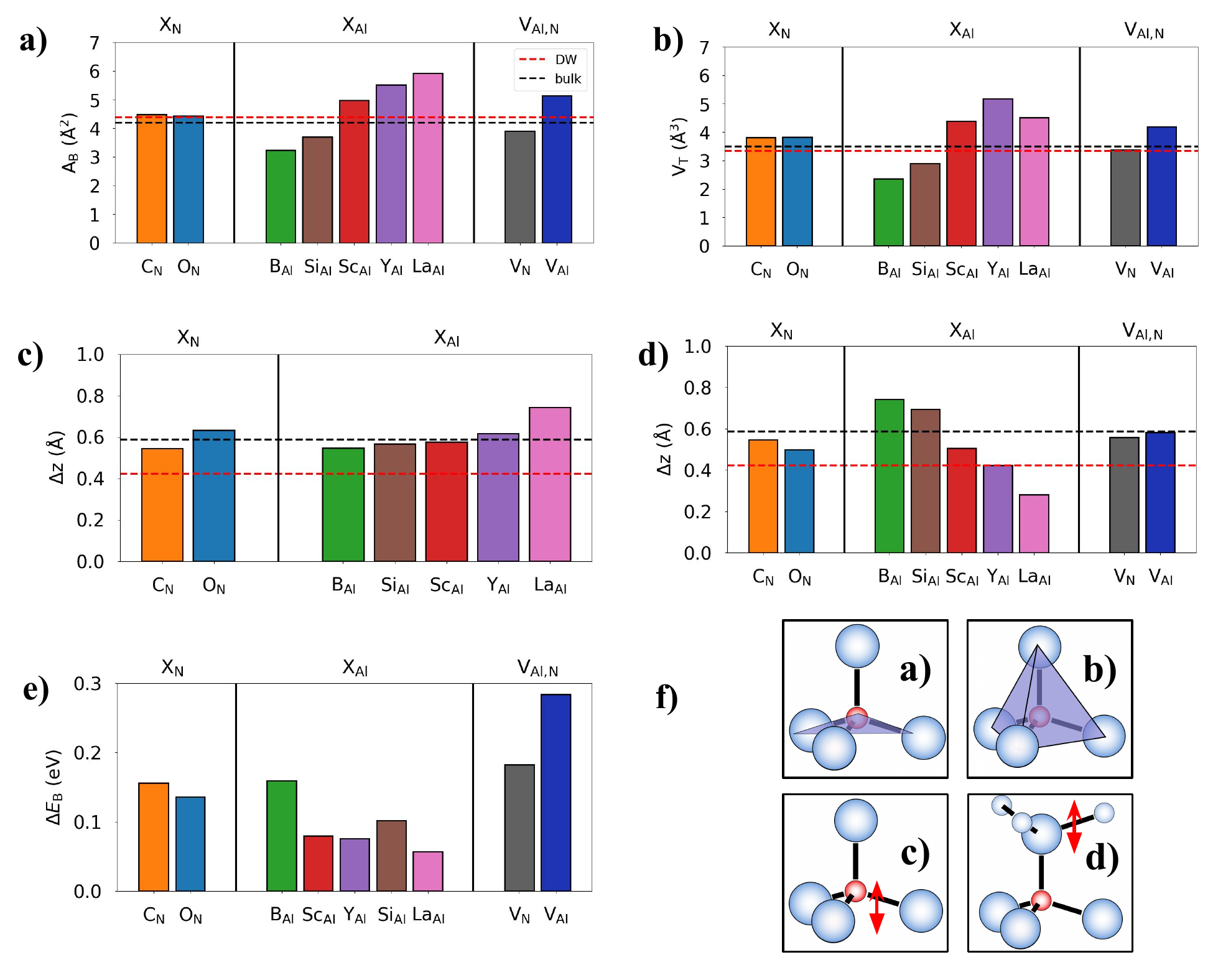}
    \caption{a) Basal area $A_\textrm{B}$, b) tetrahedral volume $V_\textrm{T}$, c) on-site and d) above-site first-neighbor $\Delta z$ parameters and e) energy difference $\Delta E_\textrm{B}$, calculated for the different neutral defects in bulk AlN. f) Schematic representation of each structural descriptors.}
    \label{fig:placeholder}
\end{figure}

\begin{figure}[h]
    \centering
    \includegraphics[width=0.95\linewidth]{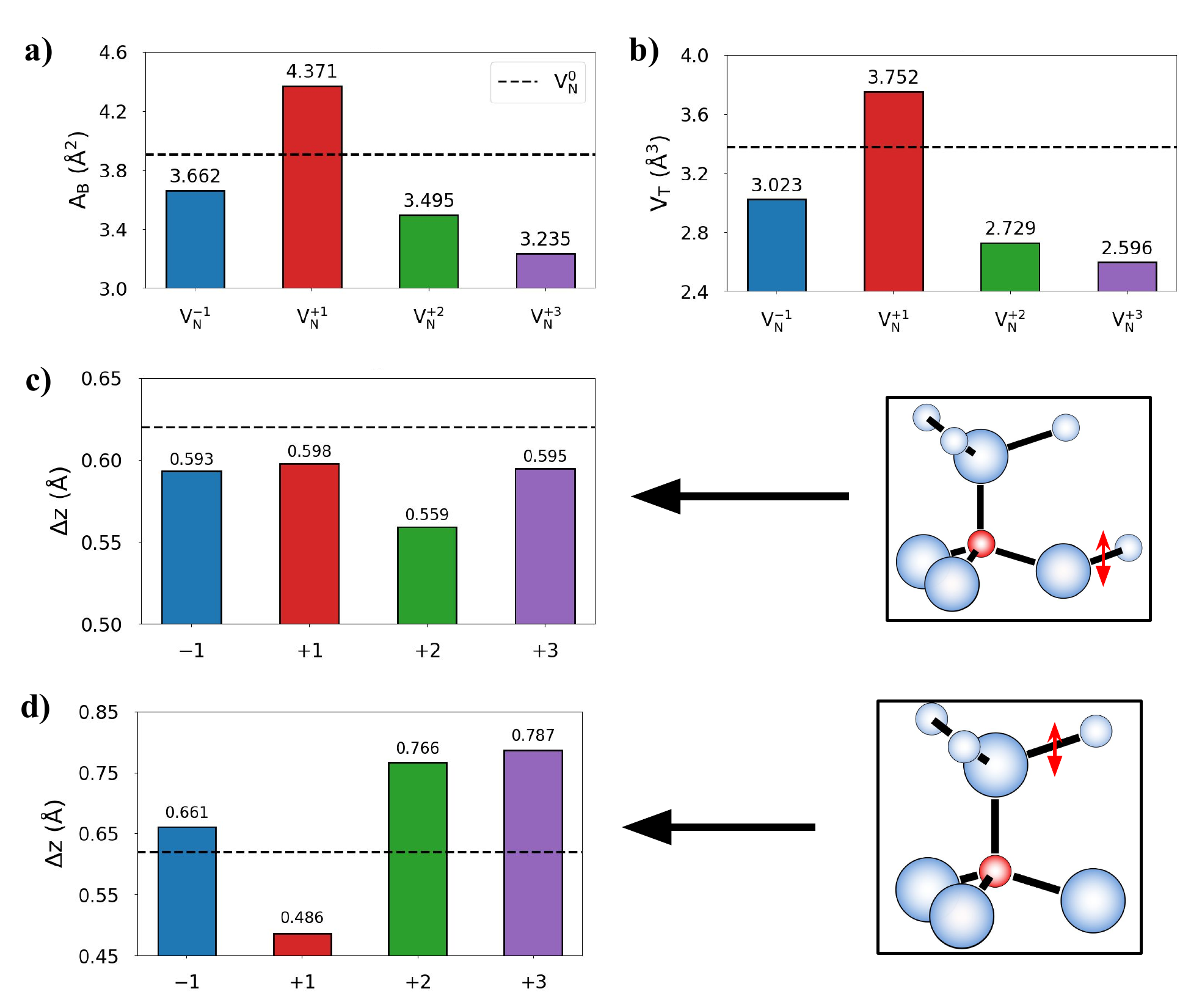}
p    \caption{a) Basal area $A_\textrm{B}$, b) tetrahedral volume $V_\textrm{T}$, c) in-plane first-neighbor, according to the schematic representation of the structural descriptor on the right and d) out-of-plane first-neighbor $\Delta z$ parameters calculated for a charged nitrogen vacancy V$_\textrm{N}^{q}$ in bulk AlN.    } 
    \label{fig:placeholder}
\end{figure}

\newpage

\subsubsection{Landau model to interpret the energy stability of the defects near the Domain Wall:}

According to the Landau model proposed in Ref.~\cite{Paillard:2017}, the energy of a unit cell of a ferroelectric compound can be expressed as:

\begin{equation}
\label{eq:Eq1}
E_\textrm{bulk} = \gamma P^2+\frac{\beta}{4}P^4+c\eta^2
\end{equation}
with $\gamma = -\kappa\eta+\frac{\alpha}{2}$. $P$ is the spontaneous electric polarization in the system and $\eta = (c-c_\textrm{cubic})/c_\textrm{cubic}$ corresponds to the strain along the polarization axis, with $c$ and $c_\textrm{cubic}$ the lattice parameters along the polarization direction in the ferroelectric and in the non-polar cubic phases. $\alpha$, $\beta$ and $\kappa$ are coefficients of the energy expansion. 

Let us consider a structure formed by a ferroelectric domain of $N_\textrm{B}$ unit cells with a spontaneous polarization $P_\textrm{B}$ and a polar-neutral domain wall (DW) of $N_\textrm{D}$ unit cells and polarization $P_\textrm{D}$. The interface between the DW and the polar domain is the volume containing $N_\textrm{DB}$ unit cells, in which the DW has a perturbative action and an average polarization $P_\textrm{DB}$. In the presence of a point defect $i$ of charge state $q$ at the center of the unpertubed polar domain, according to Eq.~\ref{eq:Eq1}, the energy of this system becomes:

\begin{align}
		\begin{split}
\label{eq:Eq2}
E_\textrm{Bulk}^i(q) = \Delta E^i_\textrm{bond}(q)&+ (N_\textrm{B}+N_\textrm{D}+N_\textrm{DB})c\eta^2 \\ &+ [N_\textrm{B}-N_\textrm{eff}(i,q)] [\gamma P_\textrm{B}^2+\frac{\beta}{4}P_\textrm{B}^4] \\ &+ N_\textrm{DB} [\gamma P_\textrm{DB}^2+\frac{\beta}{4}P_\textrm{DB}^4] \\ &+ N_\textrm{eff}(i,q)][\gamma (P_\textrm{B}+\Delta P_\textrm{B})^2 +\frac{\beta}{4}(P_\textrm{B}+\Delta P_\textrm{B})^4]
\end{split}
\end{align}

In this equation, the term $\Delta E^i_\textrm{bond}(q)$ is the energy cost for possible breakings of chemical bonds. We thus consider that the presence of the defect induce a local variation of $P$, namely $\Delta P$, in $N_\textrm{eff}(i,q)$ unit cells around the defect $i$, the polarization in the remaining $[N_\textrm{B}-N_\textrm{eff}(i,q)]$ unit cells being not affected. 

If the same defect is now located at the domain wall, we can consider that it will modify the electric polarization of in the vicinity of the DW and the energy is modified in such a way that:

\begin{align}
		\begin{split}
\label{eq:Eq3}
E_\textrm{DW}^i(q) = \Delta E^i_\textrm{bond}(q)&+ (N_\textrm{B}+N_\textrm{D}+N_\textrm{DB})c\eta^2 \\ &+ N_\textrm{B} [\gamma P_\textrm{B}^2+\frac{\beta}{4}P_\textrm{B}^4] \\ &+ [N_\textrm{DB}-N_\textrm{eff}'(i,q)] [\gamma P_\textrm{DB}^2+\frac{\beta}{4}P_\textrm{DB}^4] \\ &+ N_\textrm{eff}'(i,q)][\gamma (P_\textrm{DB}+\Delta P_\textrm{DB})^2 +\frac{\beta}{4}(P_\textrm{DB}+\Delta P_\textrm{DB})^4]
    \end{split}
\end{align}

We omit the fourth-order terms, the terms proportional to $(\Delta P^{2})$ and we do the approximation that $N_\textrm{eff}(i,q)=N_\textrm{eff}'(i,q)$, we can obtain an expression for the energy difference between the two positions of the point defect:
\begin{equation}
\label{eq:Eq4}
\Delta E_\textrm{B} = E_{[\textrm{DW}]}^i(q)-E_\textrm{Bulk}^i(q) \simeq 2\gamma N_\textrm{eff}(P_\textrm{DB}\Delta P_\textrm{DB}-P_\textrm{B}\Delta P_\textrm{B})
\end{equation}
%with $A=(N_\textrm{B}+N_\textrm{eff})(P_\textrm{DB}^2-P_\textrm{B}^2)$

\newpage
\subsubsection{Relative thermodynamic stability \textit{vs} charge state $q$:}
\begin{figure}[htb!]
    \centering
    \includegraphics[width=0.9\linewidth]{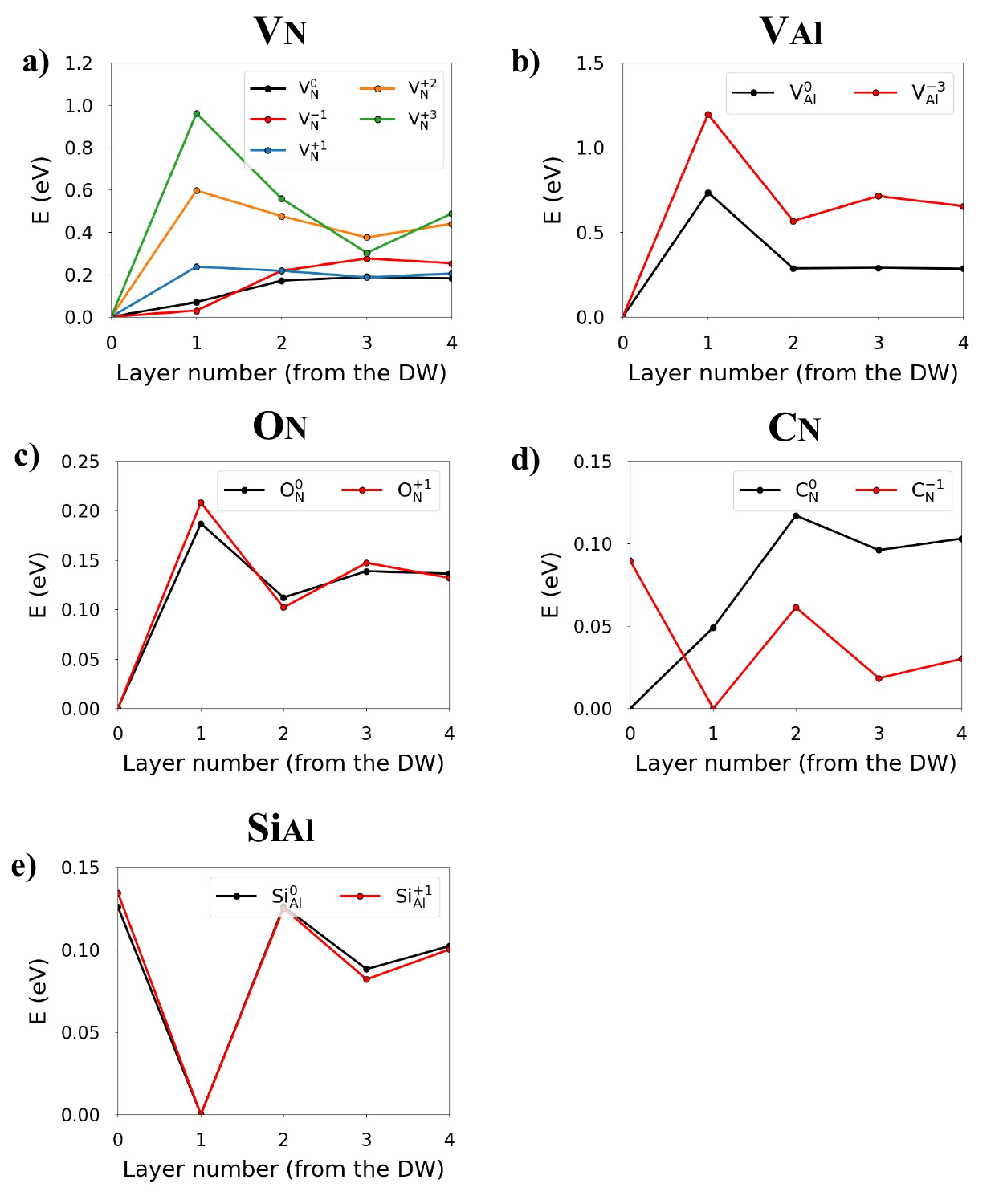}
    \caption{Comparison of energetic stability of the point defect that can be stabilized with different charge states $q$, stable according to Ref.~\cite{Lee:2024,Osetsky:2022,Aleksandrov:2020}.}
    \label{fig:placeholder}
\end{figure}

\newpage

\subsection{Migration of the vertical domain wall}

\subsubsection{without point defects:}

\begin{figure}[h]
    \centering
    \includegraphics[width=0.8\linewidth]{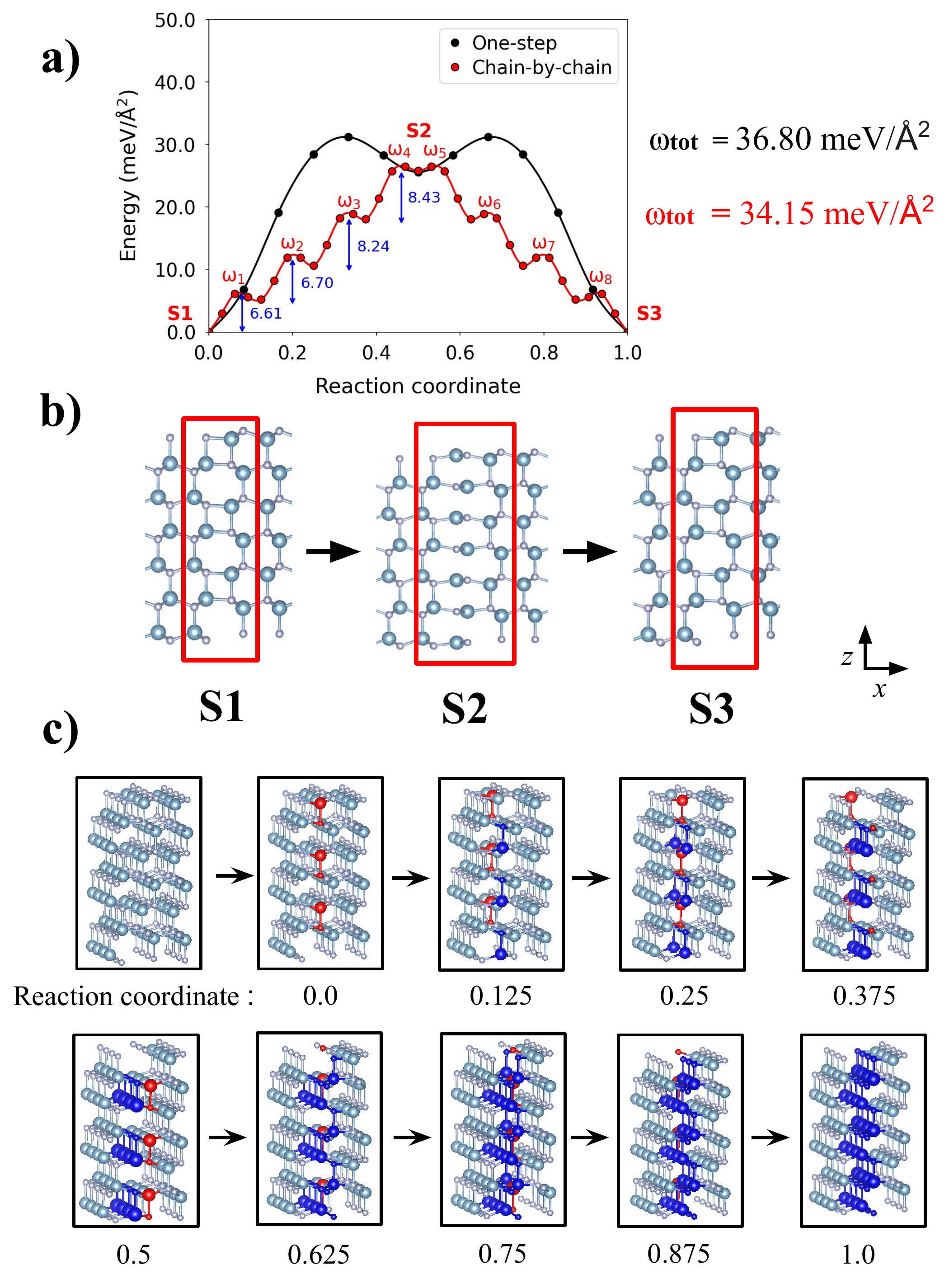}
    \caption{a) Comparison of the total-energy variations during the displacement of one DW by a distance of one unit cell along the $[10\bar{1}0]$ direction, as calculated for the two mechanisms denoted as ``one-step'' and ``chain-by-chain'' displacements. (Meta)stable atomic structures spanned during the (b) one-step or (c) chain-by-chain displacement.}
    \label{fig:figS5}
\end{figure}

As shown in Fig.~\ref{fig:figS5}, we studied the displacement of the domain wall for the supercell without any point defects. We investigated two different types of displacement mechanisms, later designed as ``one-step'' and ``chain-by-chain''. In the one-step displacement, all atoms in the DW are displaced at the same time whereas for chain-by-chain the switching occurs for all atoms of each [0001] atomic column at a time. In our supercell, an atomic column consists of 6 atoms aligned along the $+c$ direction. 

\newpage

\subsubsection{with point defects:}

\begin{figure}[htb!]
    \centering
    \includegraphics[width=0.9\linewidth]{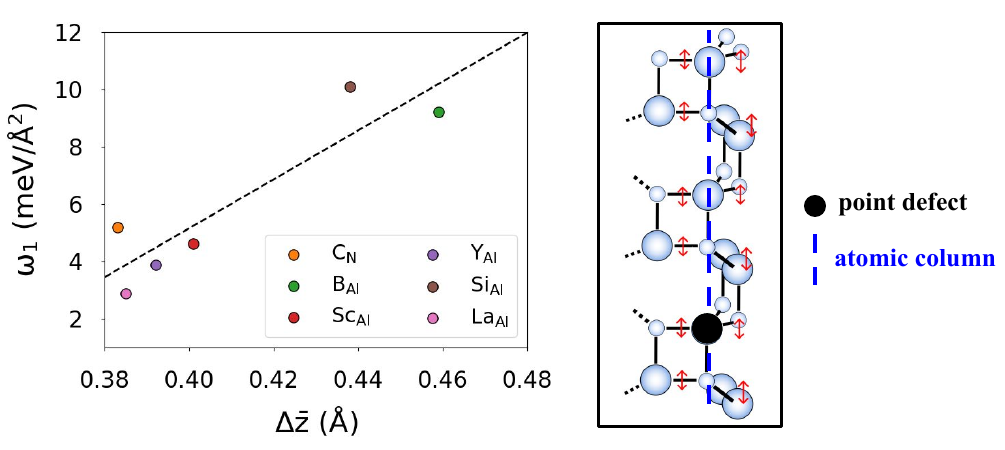}
    \caption{ Barrier energy $\mathrm{\omega_1}$ for switching the first [0001] atomic column (where the point defect is located) vs the average cation-anion rumpling $\Delta \bar{z}$.  }
    \label{fig:w1_vs_delta_z}
\end{figure}

%\remi{pas sûr si Delta z est bien défini dans cette figure}

\begin{figure}[h]
    \centering
    \includegraphics[width=1.0\linewidth]{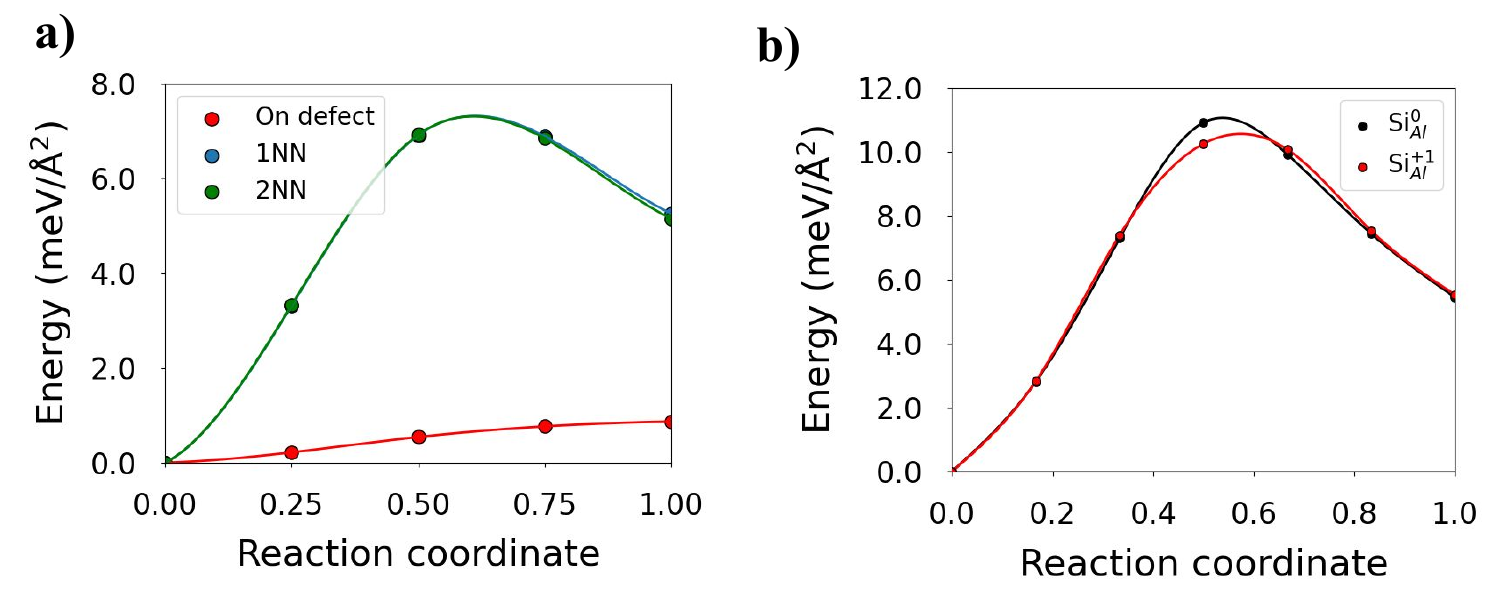}
    \caption{a) Evolution of the total energy during the switching of a first the atomic column of a vertical inversion domain wall containing one O atom substitution. The red curve corresponds to the configuration when the first column contains the O$_\textrm{N}$ substitution and we can see that in this case, the switching does not end to a metastable structure. Instead, if we switch the column, first- or second-neighbor (1NN/blue or 2NN/green) of the defect, an energy barrier and a local minimum appear. b) First-column switching energy profile for a Si substitution and two different charge states $q = 0$ (black) and $q = +1$ (red).}
    \label{fig:placeholder}
\end{figure}

\begin{figure}[htb!]
    \centering
    \includegraphics[width=0.9\linewidth]{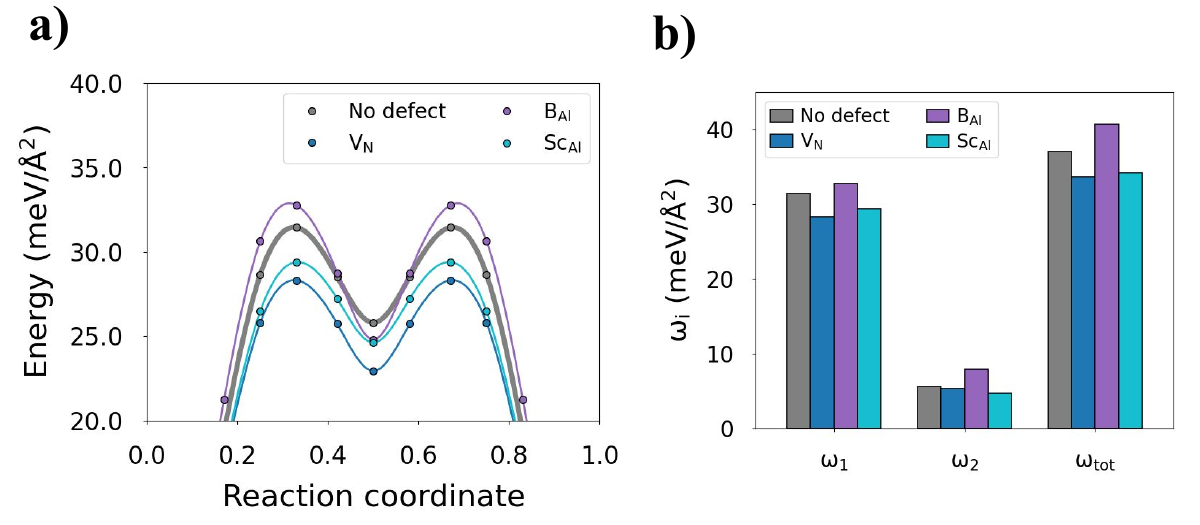}
    \caption{a) NEB energy profile for different point defects ( V$_\textrm{N}$,B$_\textrm{Al}$, Sc$_\textrm{Al}$) compared to pristine AlN, only calculated for the collective-switching mechanism. b) Histogram showing the two displacement steps $\mathrm{\omega_{1,2}}$ and the total sum   $\mathrm{\omega_{tot}}$ for the four different configurations.}
    \label{fig:figS8}
\end{figure}
Because calculating the full chain-by-chain reaction path to displace the DW for every defect was too computationally demanding, we choose to compute the variation of energy barrier for a coherent displacement of the DW for the most significant defects, this in order to have a clearer idea of the pinning effect. As it can be seen in Fig.~\ref{fig:figS8}, with this displacement mechanism, the structural changes go through one metastable state and two energy barriers have to be overcome. For a nondefective DW, the total barrier is calculated to $\omega_\textrm{tot} = 37.09$~meV~\AA$^{-2}$. It is reduced by 9.1\% and 7.8\% with a V$_\textrm{N}$ or Sc$_\textrm{Al}$ defect, respectively, while B$_\textrm{Al}$ substitution leads to an increase of $\omega_\textrm{tot}$ by 9.7\%.  Again, on the contrary to other defects, it is found with this mechanism that the presence of a O$_\textrm{N}$ substitution does not allow for the stabilization of an intermediate metastable state, and the switching occurs through a single barrier of $\omega = 59.13$~meV~\AA$^{-2}$. 

\newpage

\subsection{Densities of states near a point defect at the domain wall}

\begin{figure}[htb!]
    \centering
    \includegraphics[width=0.75\linewidth]{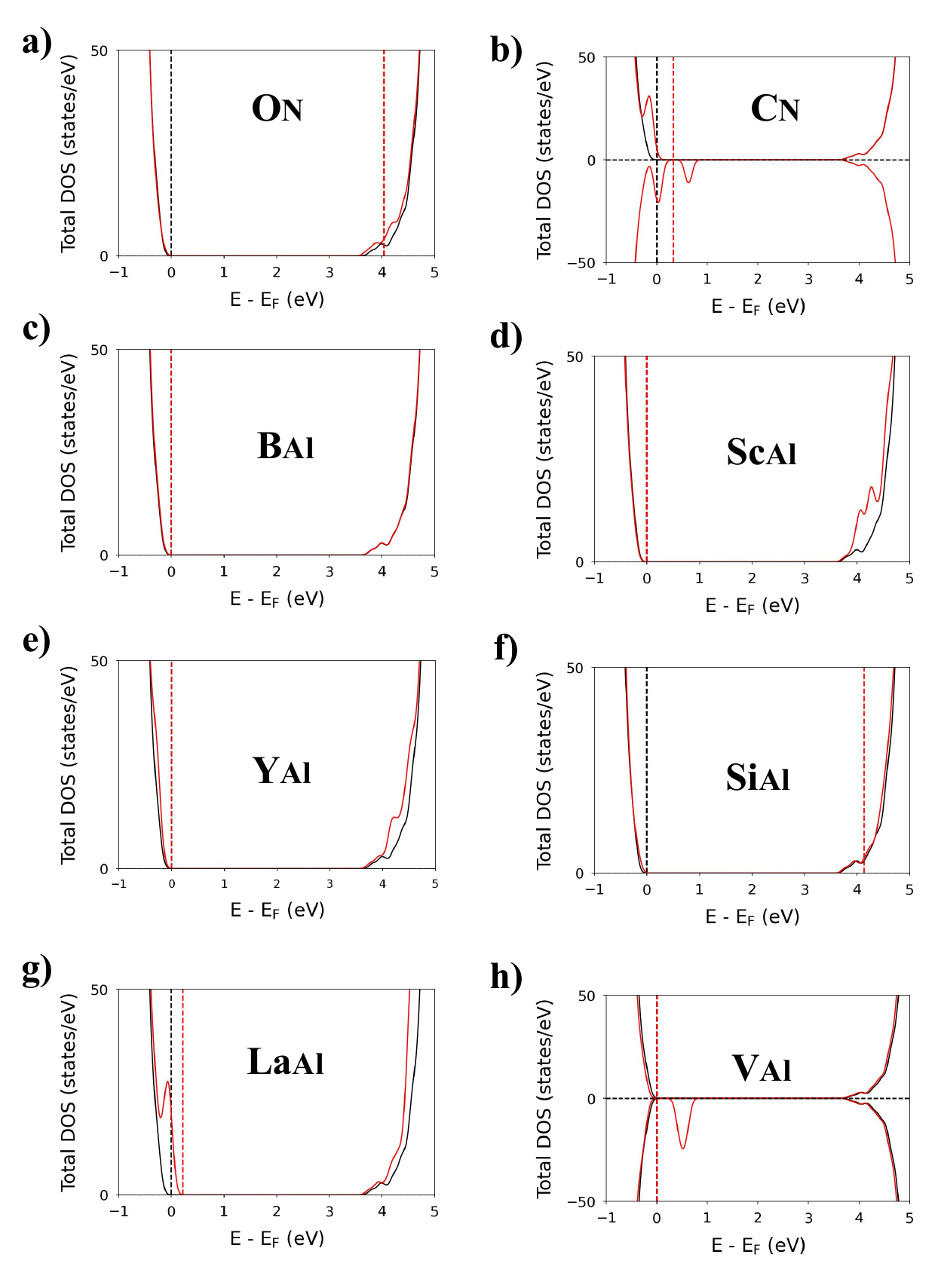}
    \caption{Densities of states calculated calculated in AlN for charge-neutral point defects located in their more stable position near the domain wall. The DOS of the defective structures (red curves) are compared with those of the undefective bulk AlN (black curves) to highlight the formation of defect states and the modification of the band-gap width.}
    \label{fig:placeholder}
\end{figure}

\begin{figure}
    \centering
    \includegraphics[width=1.0\linewidth]{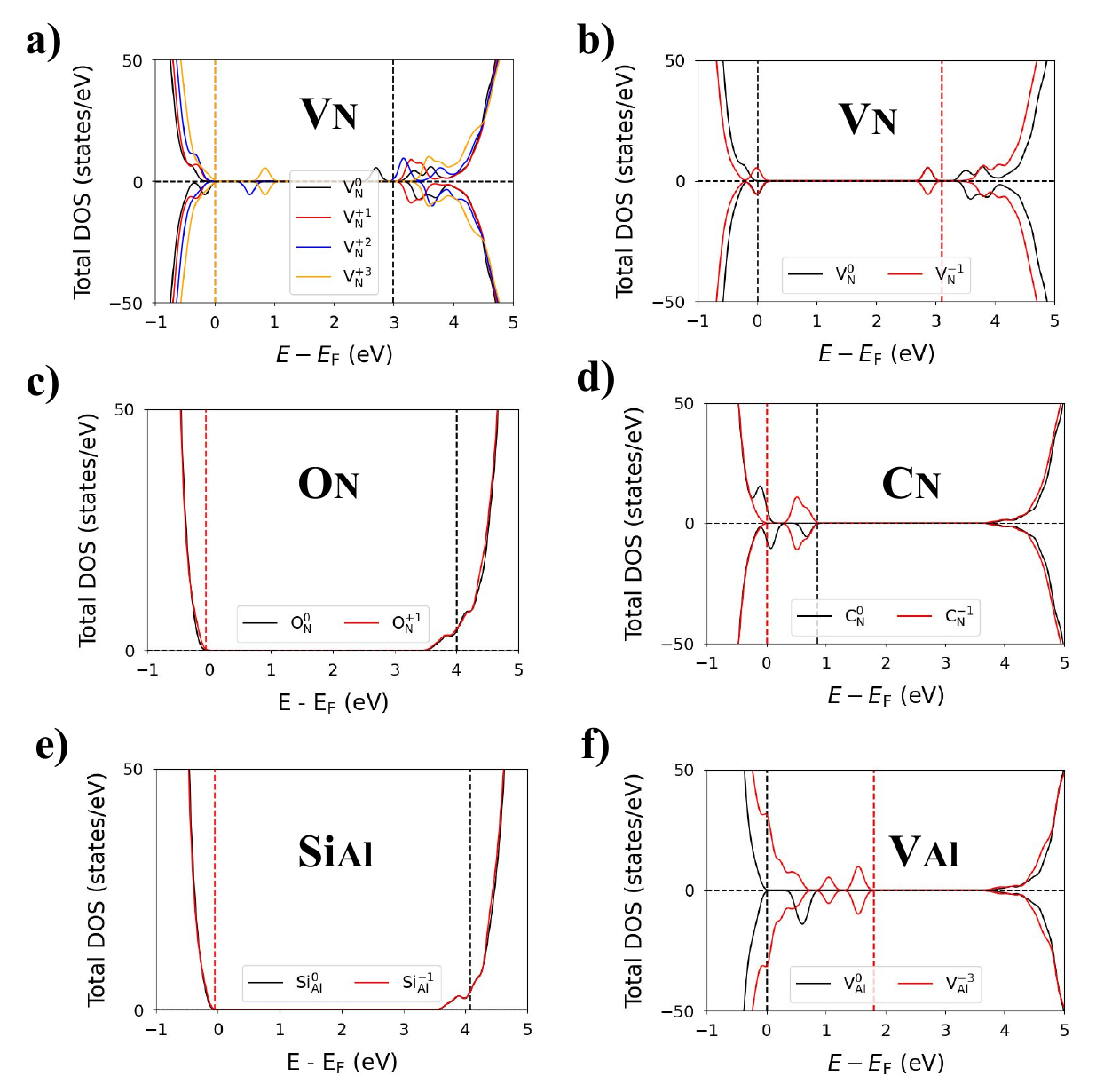}
    \caption{Densities of states calculated calculated in AlN for point defects located in their more stable position near the domain wall and for different charge states $q$.}
    \label{fig:placeholder}
\end{figure}

\newpage

\newpage

\bibliography{biblio_SI}

\end{document}